\documentclass[aps,prl,twocolumn]{revtex4}

\usepackage{graphicx}
\usepackage{dcolumn}
\usepackage{amsmath}

\begin{document}

\title{Magneto-electric excitations in multiferroic TbMnO$_3$ by Raman scattering}

\author{P. Rovillain$^1$}
\author{M. Cazayous$^1$}
\author{Y. Gallais$^1$}
\author{A. Sacuto$^1$}
\author{M-A. Measson$^1$}
\author{H. Sakata$^2$}
\affiliation{$^1$Laboratoire Mat\'eriaux et Ph\'enom\`enes Quantiques (UMR 7162 CNRS), Universit\'e Paris Diderot-Paris 7, 75205 Paris cedex 13, France\\
$^2$Department of Physics, Tokyo University of Science, 1-3 Kagurazaka Shinjyuku-ku Tokyo, Japan 162-8601}



\date{\today}
     
\begin{abstract}
Low energy excitations in the multiferroic material TbMnO$_3$ have been investigated by Raman spectroscopy. 
Our observations reveal the existence of two peaks at 30~cm$^{-1}$ and 60~cm$^{-1}$. They are observed in the cycloidal phase below the Curie temperature but not in the sinusoidal phase, suggesting their magnetoelectric origin. While the peak energies coincide with the frequencies of electromagnons measured previously by transmission spectroscopy, they show surprisingly different selection rules, with the 30~cm$^{-1}$ excitation enhanced by the electric field of light along the spontaneous polarization. The origins of the modes are discussed under Raman and infrared selection rules considerations. 
\end{abstract}


\maketitle

Multiferroics have both ferroelectricity and magnetism. For some of these materials, the magnetoelectric coupling is especially strong and has attracted much attention for new spin-based device applications \cite{Bea}. Substantial efforts have been dedicated to the reseach on the origin of the close coupling between the magnetic and electric orders. TbMnO$_3$ is one of the most intensively studied magnetoelectric manganite among the frustated magnets. The ferroelectricity in TbMnO$_3$ appears to be induced by an inverse Dzyaloshinski-Moriya interaction \cite{Katsura, Sergienko}, even if the microscopic mechanism remains under debate \cite{Sergienko2}. The strength of the magnetoelectric coupling gives rise to dynamical effects like electromagnons, magnons with an electric dipole activity predicted by Baryachtar and Chupis \cite{Baryakhtar}.
Such excitations have been observed by far infrared transmission \cite{Pimenov, Takahashi, Sushkov, Valdes, Lee, Pimenov2}. This spectroscopy detects electromagnons for electric field of light {\bf E} parallel to the $a$ axis of the crystal at around 2.5~meV (20~cm$^{-1}$) and 7.5~meV (60~cm$^{-1}$). Inelastic neutron measurements detect magnetic excitations at the same energies along the same crystallographic direction \cite{Sneff}. 
From a theoretical point of view, Katsura {\it et al.} have proposed a model based on spins current to describe the electromagnons \cite{Katsura2}. 
This model can be regarded as an inverse Dzyaloshinskii-Moriya effect and predicts the observation of one electromagnon (25~cm$^{-1}$) with selection rule {\bf E}~$\parallel$~a-axis perpendicular to the bc plane in the spiral phase. 
Recent approaches based on indirect Heisenberg exchange \cite{Sushkov2} and on cross-coupling between magnetostriction and spin-orbit interactions \cite{Sousa} were proposed to explain respectively one electromagnon (60~cm$^{-1}$) and the both excitations (30 and 60~cm$^{-1}$). 
Optical spectroscopies have different selection rules and the scattering processes involved in each spectroscopy should be differently sensitive to the electric dipole activity of the electromagnons. 
Among them, Raman scattering is an efficient probe for studying both magnetic (magnons) and ferroelectric (phonons) excitations and their mutual coupling \cite{Cazayous, Singh, Scott1, Cazayous2}. However, up to now the electromagnon signature in TbMnO$_3$ has not been detected by Raman scattering. 

\par
Here, we investigate the magnetic excitations in TbMnO$_3$ through Raman measurements. Our study reveals magnons at 30~cm$^{-1}$ and 60~cm$^{-1}$ with the electric field of light {\bf E}~$\parallel$~a. The intensity of the magnon at 30~cm$^{-1}$ is enhanced with electric field {\bf E}~$\parallel$~c, the magnon at 60~cm$^{-1}$ disappears and a strong band is detected at 128~cm$^{-1}$. Both magnetic modes (30~cm$^{-1}$ and 60~cm$^{-1}$) are only observed in the ferroelectric phase (cycloidal phase) which points out their electric-dipole activity. The band at 128~cm$^{-1}$ presents the same frequency shift as a function of the temperature as the 113~cm$^{-1}$ polar phonon. In a magnon-phonon scattering scenario, this indicates that the band at 128~cm$^{-1}$ comes from a second order scattering process involving the 30~cm$^{-1}$ magnon and the 113~cm$^{-1}$ phonon. 

\begin{figure}
\includegraphics*[width=8cm]{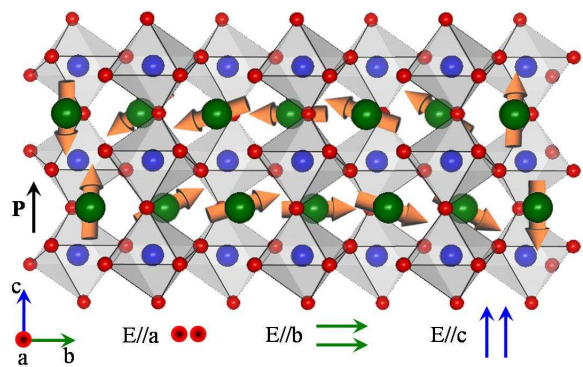}\\
\caption{\label{Figure1} 	 
Structure of the orthorhombic TbMnO$_3$ crystal whith the cycloid ordering of the Mn spins in the ferroelectric phase below T$_C$. The spins rotate in the $bc$ plane around the $a$ axis and propagate along the $b$ direction. Parallel polarizations of the incident and scattered electric fields {\bf E} along the $a$, $b$ and $c$ axes.}
\end{figure}

\par
TbMnO$_3$ single crystals were grown using the floating zone method and crystallize in the orthorhombic symmetry of space group P{\it bnm} \cite{Sakata}. 
Below the N\'eel temperature $T_N=42$~K the Mn magnetic moments order antiferromagnetically in an incommensurate sinusoidal wave with a modulation vector along the $b$ axis (sinusoidal phase). At still lower temperature $T_C=28$~K, the spin wave modulation continuously transforms into a cycloid (cycloidal phase) with spins confined to the $bc$ plane (Fig.~\ref{Figure1}). This transition is associated with the appearance of a spontaneous electric polarization $P$ along the $c$ axis.
In this work, two samples with $ac$ and $bc$ planes have been investigated.

\par
We have performed Raman measurements in a backscattering geometry with a triple spectrometer Jobin Yvon T64000 using the 568~nm excitation line from a Ar$^+$-Kr$^+$ mixed gas laser. Tiny signals have been obtained with other laser wavelengths. The high rejection rate of the spectrometer allows us to detect the magnons at frequencies below 100~cm$^{-1}$. The temperature dependences have been performed using a ARS closed cycle He cryostat. Figure~\ref{Figure1} shows the two configurations of light polarizations used. Incident and scattered lights are polarized along the same crystallographic axis. 

\begin{figure}
\includegraphics*[width=8cm]{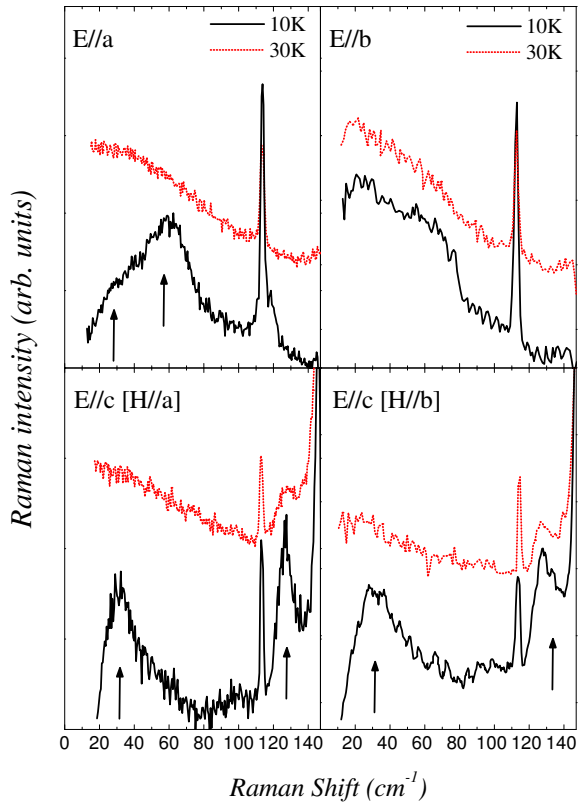}\\
\caption{\label{Figure2} 	 
Raman response measured in the cycloidal (10~K) and sinusoidal phase (30~K) using different configurations for the electric {\bf E} and magnetic {\bf H} fields of light. Arrows show the both magnon modes and the band at 30, 60 and 128~cm$^{-1}$, respectively.}
\end{figure}

\par
Figure~\ref{Figure2} shows the Raman response with different light polarizations in the cycloidal (below T$_C$) and sinusoidal phases (below T$_N$). 
({\bf E}, {\bf H}) are the electric and magnetic fields of the light, respectively. 
In {\bf E}~$\parallel$~a, a strong peak is observed at 60~cm$^{-1}$ and a shoulder at 30~cm$^{-1}$ (10~K) and both disappear at 30~K. 
The signature of the both peaks seems to be present in {\bf E}~$\parallel$~b but with a negligible intensity. 
Using {\bf E}~$\parallel$~c ({\bf H}~$\parallel$~a or {\bf H}~$\parallel$~b) the peak at 60~cm$^{-1}$ disappears, the peak at 30~cm$^{-1}$ grows up with a band at 128~cm$^{-1}$, both with the same intensities in the two configurations. 
No phonon is expected under 100 cm$^{-1}$. These peaks can be thus attributed to magnon modes. However, the origin of the two magnon modes is not obvious and is discussed below. 

\begin{figure}
\includegraphics*[width=7.5cm]{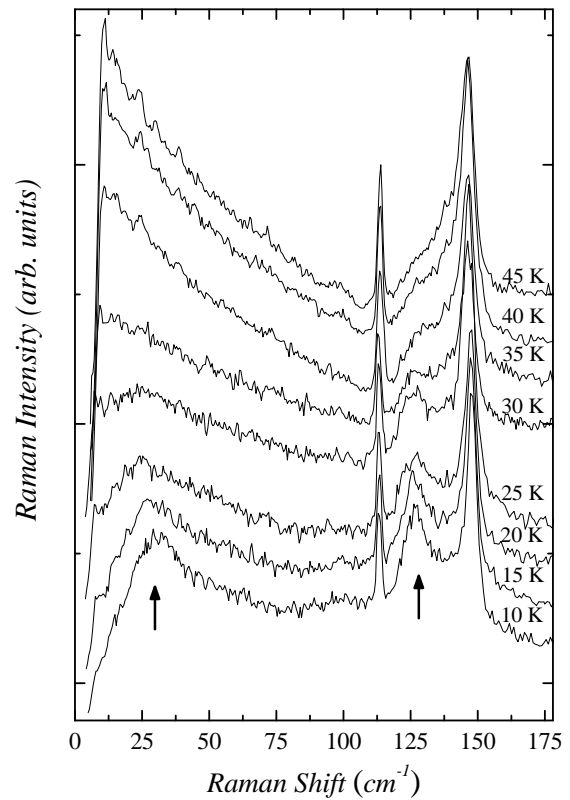}\\
\caption{\label{Figure3} 
Temperature-dependent Raman spectra in {\bf E}~$\parallel$~c between 10 and 45~K. Observation of two pics at 30 and 128~cm$^{-1}$ and two phonons at 113 and 147~cm$^{-1}$.}
\end{figure}

\begin{figure}
\includegraphics*[width=8.5cm]{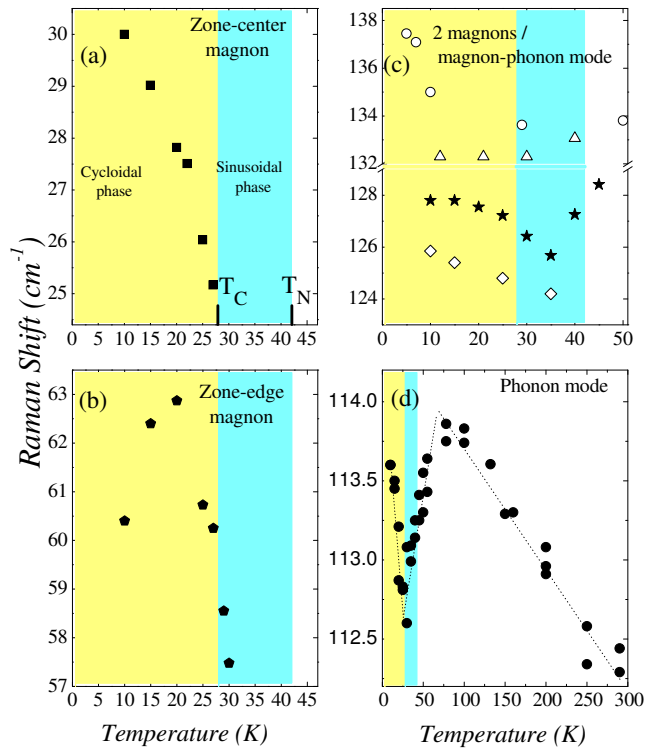}\\
\caption{\label{Figure4}
Our measurements (full symbol) : frequency of the (a) 30~cm$^{-1}$ (square), (b) 60~cm$^{-1}$ magnons modes (hexagon), of the (c) 128~cm$^{-1}$ band (star) and of the (d) 113~cm$^{-1}$ phonon mode (circle) as a function of the temperature. (c) Open symbols :  far infrared data of Schmidt {\it et al.} (open circle) \cite{Schmidt} and Takahashi {\it et al.} (open triangle) \cite{Takahashi}, and Raman measurements of Barath {\it et al.} (open diamond) \cite{Barath}. The color zones define the cycloidal phase from 0~K to 28~K and the sinusoidal phase from 28~K to 42~K. Lines are guide to the eye.} 
\end{figure}

Figure~\ref{Figure3} shows the temperature dependence of Raman spectra in the range 0-175~cm$^{-1}$ from 10 to 45~K with {\bf E}~$\parallel$~c. The magnon mode at 30~cm$^{-1}$ clearly disappears upon is entering in the sinusoïdal phase (T>30~K) whereas the band at 128~cm$^{-1}$ is still observed in the collinear sinusoidal phase before finally vanishing at the N\'eel temperature ($T_N=42$~K). 
Figure~\ref{Figure4}(a) shows quantitatively that the frequency of the magnon mode at 30~cm$^{-1}$ decreases down to the end of the cycloidal phase. The frequency of 60~cm$^{-1}$ magnon (Fig.~\ref{Figure4}(b)) first increases from 10~K up to 20~K before decreasing until 30~K. These both magnetic excitations are only detected in the cycloidal phase whereas they should exist in the sinusoidal phase until the N\'eel temperature $T_N=42$~K as expected for ordinary magnetic excitations. Our data show that both excitations have not a pure magnetic activity and are intimetly related to the cycloidal phase below T$_C$. 

Let us focus on the origin of the observed magnetic peaks and on the Raman polarization selection rules.

The peak at 60~cm$^{-1}$ is assigned to a zone-edge magnon with an energy close to the zone-edge energy \cite{Lee, Valdes}. 
 Previously measured at the same energy and with the same polarization {\bf E}~$\parallel$~a by far infrared transmission spectroscopy, this magnetic excitation has been identified as an electromagnon \cite{Pimenov, Takahashi}. Raman measurements show that this magnetic excitation exists only in the cycloidal phase underlying the polar caracter of the zone-edge magnon. 
 Optical spectroscopies like Raman scattering probe dispersion branches close to the zero wave vector.
The activation of 60~cm$^{-1}$ zone-edge magnon can be explained by the alternation of the Heisenberg exchange interaction along $b$ axis \cite{Valdes} or by the coupling of this mode with the spontaneous polarization through the dynamical magnetoelectric field \cite{Sousa}.
  
The origin of the 30~cm$^{-1}$ peak is more tricky. 
It might be assigned to zone-center magnon mode.
A broad peak has been already reported by infrared between 20~cm$^{-1}$ and 25~cm$^{-1}$ and has been assigned to electromagnon \cite{Pimenov, Takahashi, Pimenov2}. Based on neutron measurements, the peak observed by infrared corresponds to a propagating mode of the spins out of the cycloidal plane \cite{Sneff}. However, infrared measurements show only this peak with {\bf E}~$\parallel$~a whereas this peak is enhanced with {\bf E}~$\parallel$~c in Raman scattering. 
Raman scattering is induced by the electric field {\bf E} of the light irrespective to the polarization direction of {\bf H}.
No significant Raman signal has been measured using light cross polarizations. This result is unexpected refering to the Fleury and Loudon approach  showing that the one-magnon response exists for cross configuration \cite{Loudon}. This unexpected selection rule might be interpreted as the electric-dipole activity of the magnon mode in TbMnO$_3$. 
Raman scattering is very different from IR experiments whose electromagnon peaks arise directly from electric dipole activity of magnons. 
The selections rules (the polarization of the electric field of the light) involved in the Raman scattering process might be not directly connected to the polarization that activates the electromagnon. Recently, A. Cano  shows that the magnetoelectric response giving rise to the electromagnon features in optical experiments not always can be reduced to an effective electric permittivity \cite{Cano}. This might induce discrepancy between the selection rules of different experimental techniques. We can notice that the 30~cm$^{-1}$ mode measured by Raman scattering has a higher frequency compare to the electromagnons measured by infrared spectrosopies (20-25~cm$^{-1}$). These optical spectroscopies might be differently sensitive to the hybridization degree of electromagnons. The observation of this electromagnon at lower energy in infrared spectroscopy would suggest that infrared spectroscopy is more affected by the polar activity of the electromagnons than the Raman one. 
The Dzyaloshinskii-Moriya interaction has been proposed to explain the electromagnon at 30~cm$^{-1}$ with a polar activity predicted for the electric field of light along the $a$ axis \cite{Katsura2}. A recent model based on cross-coupling between magnetostriction and spin-orbit interactions can explain the both peak at 30~cm$^{-1}$ and 60~cm$^{-1}$ \cite{Sousa}. 
In this model the 30~cm$^{-1}$ is not connected to zone-center magnon mode but corresponds to an excitation combining the zone-edge magnon wavevector and twice the cycloid wavevector. 

\par
In Fig.~\ref{Figure3} the phonon modes at 113~cm$^{-1}$ and 148~cm$^{-1}$ correspond to A$_g$-symmetry modes associated to displacements of the Tb$^{3+}$ ions \cite{Venugopalan,Martin}. The mode at 113~cm$^{-1}$ is measured with Raman scattering in the two polarizations {\bf E}~$\parallel$~a and {\bf E}~$\parallel$~c whereas the mode at 148~cm$^{-1}$ is only present in the polarization {\bf E}~$\parallel$~c.
The behaviour of the lowest phonon mode at 113~cm$^{-1}$ is unusual (Fig.~\ref{Figure4}(d)) with a  sharp frequency  decreasing in the cycloidal phase followed by an increase up to 75~K before the usual frequency decrease at higher temperature due to the thermal expansion of the lattice. 

\par
The detection of phonon anomalies related to ferroelectricity is a quest to determine the microscopic mechanism involved and to explain how the spontaneous polarization appears \cite{Katsura}. Barath et al. have observed the evolution of the phonon mode at 147~cm$^{-1}$ under a magnetic field \cite{Barath}. More recently, no phonon anomalies were found below 400~cm$^{-1}$ by X-ray scattering suggesting a non conventional displacive ferroelectric transition in TbMnO$_3$ \cite{Kajimoto}. Here, we clearly observe an anomaly in a $c$-polarized phonon frequency at 113~cm$^{-1}$ across T$_C$ (Fig.~\ref{Figure4}(d)). The small frequency shift ($\Delta\omega$~=~1~$\pm$~0.15~cm$^{-1}$) of this mode shows that the coupling between the spontaneous polarization and the lattice is weak and confirms the magnetic origin of the ferroelectricity.

\par
In Fig.~\ref{Figure4}(c), the frequency of the band at 128~cm$^{-1}$ is shown with previously reported far infrared and Raman measurements.
It decreases from 10 to 35~K and increases up to 45~K. The measurements performed by Schmidt {\it et al.} \cite{Schmidt} and Barath {\it et al.} \cite{Barath} present the same decrease up to 35~K. The data of Takahashi {\it et al.} show a small discrepancy \cite{Takahashi}. 

The origin of this band has been already discussed and interpreted as a magnon, a two-magnon scattering or a magnon-phonon process \cite {Takahashi, Lee, Schmidt}. 
As already mentionned this band disappears at the N\'eel temperature which underlines its magnetic character. 
This band can not be associated with a one magnon procces because the zone-edge for magnetic excitations is around 60~cm$^{-1}$.
   
First, the band at 128~cm$^{-1}$ can be explained by the two-magnon scattering process i.e. twice the magnon energy at 60~cm$^{-1}$ \cite{Takahashi}.  The zone-edge magnon at 60~cm$^{-1}$ disappears at T$_C$ whereas the two-magnon scattering process associated with this mode is still measured in the cycloidal phase up to T$_N$. This points out the mixed character of the zone-edge magnon in the two magnon scattering picture.   

Second, we consider the one-magnon + one-phonon scenario \cite{Lockwood}. In the range 10-45~K, the temperature dependence of the 128~cm$^{-1}$ band (Fig.~\ref{Figure4}(c)) is similar to the one of the 113~cm$^{-1}$ phonon mode (Fig.~\ref{Figure4}(d)). This indicates that the band at 128~cm$^{-1}$ can arise from a magnon-phonon scattering process involving the magnon at 30~cm$^{-1}$ and the 113~cm$^{-1}$ phonon.
In this scenario, our data show a strong coupling between the lowest optical phonon and the 30~cm$^{-1}$ magnon mode. 
The two scenarios discussed here advocate in favour of the electric-dipole ac tivity of the Raman magnon modes observed at 30 and 60~cm$^{-1}$.

\par
In conclusion, our Raman observations reveal two magnetic excitations at 30~cm$^{-1}$ and 60~cm$^{-1}$ with light polarization {\bf E}~$\parallel$~a.  Surprisingly, our measurements show that the 30~cm$^{-1}$ mode is enhanced with a light electric field along the spontaneous polarization ($c$ axis). Both modes are only present in the cycloidal phase below T$_C$. The mode at 60~cm$^{-1}$ is interpreted as the zone-edge magnon-phonon hybridization with the phonon part describing the electric polarization parallel to $a$. The Raman selection rules for 30~cm$^{-1}$ excitation show its complex origin. 
Finally, we show the intimate relationship between the optical 113~cm$^{-1}$ phonon mode and the mode at 128~cm$^{-1}$. 

\par
\par
The authors would like to thank R. de Sousa for helpful discussions and for a critical reading of the manuscript.

\end{document}